\documentclass[11pt,a4paper]{article}

\usepackage[utf8]{inputenc}
\usepackage[T1]{fontenc}
\usepackage{lmodern}
\usepackage[margin=1in]{geometry}
\usepackage{amsmath,amssymb}
\usepackage{chngcntr}
\usepackage{graphicx}
\usepackage{authblk}
\usepackage{hyperref}
\usepackage[numbers,sort&compress]{natbib}
\usepackage{microtype}
\usepackage{url}

\hypersetup{
    colorlinks=true,
    linkcolor=blue,
    citecolor=blue,
    urlcolor=blue
}

\newif\ifincludesupp

\includesupptrue

\title{
PyFLI: A Python Library for Simulation, Parameter Estimation, and Benchmarking in Fluorescence Lifetime Imaging
}

\author[1,*]{Vikas Pandey}
\author[1]{Ismail Erbas}
\author[2]{Margarida Barroso}
\author[1]{Stefan Radev}
\author[1]{Xavier Intes}

\affil[1]{
Center for Modeling, Simulation and Imaging in Medicine,
Rensselaer Polytechnic Institute, USA
}

\affil[2]{
Department of Molecular and Cellular Physiology, Albany Medical College, USA
}

\affil[*]{
Corresponding authors:
\href{mailto:pandev2@rpi.edu}{pandev2@rpi.edu},
\href{mailto:erbasi@rpi.edu}{erbasi@rpi.edu}
}

\date{}

\begin{document}

\maketitle


\section{Summary}

Fluorescence lifetime imaging (FLI) measures the temporal decay of
fluorescence after excitation and provides quantitative information
about a fluorophore's local environment and molecular interactions.
Depending on the fluorophore and experimental design, lifetime can
report changes associated with pH, ion concentration, viscosity,
oxygenation, cellular metabolism, and F{\"o}rster resonance energy
transfer (FRET). These properties make FLI useful across microscopy,
biophysics, biomedical optics, and preclinical imaging, where the same
type of molecular contrast can be studied across different biological
scales.

FLI measurements, however, are acquired with instruments that record
fluorescence in different ways. Intensified charge-coupled device (ICCD)
cameras, single-photon avalanche diode (SPAD) arrays, and
time-correlated single-photon counting (TCSPC) systems differ in
temporal sampling, data organization, detector noise, instrument
response, and file format. Lifetime-estimation methods also make
different assumptions about the recorded decay, while learning-based
approaches require realistic training and validation data for which the underlying parameters are known.

PyFLI is an open-source Python framework for experimental FLI processing
and standardized FLI data generation. It imports and preprocesses
measurements from supported acquisition systems, generates labeled
synthetic data under configurable acquisition and noise conditions, and
provides complementary approaches for lifetime-parameter estimation.
These include nonlinear least-squares fitting (NLSF),
maximum-likelihood estimation (MLE), Phasor analysis, rapid lifetime
determination (RLD), Laguerre-based estimation, and an optional trained
posterior-sampling pathway for Bayesian and deep-learning inference.
CPU and GPU processing support image-scale analysis, while
reconstruction, visualization, statistical analysis, and cross-software
comparison provide common tools for evaluating results. PyFLI also
includes compressed-sensing reconstruction for single-pixel
hyperspectral FLI.

\section{Statement of need}

A fluorescence-lifetime measurement reflects both the underlying
fluorescence decay and the acquisition system used to record it. ICCD
systems commonly sample fluorescence through nanosecond time gates and
include intensifier gain noise. SPAD systems acquire gated
single-photon measurements with detector-specific counting statistics
and dead-time effects \citep{Bruschini2019}. TCSPC systems construct
photon-arrival histograms and can be affected by pile-up at high count
rates. Researchers therefore work with gated image sequences,
photon-count histograms, microscopy containers, generic numerical
arrays, separate instrument-response functions (IRFs), background
measurements, masks, and different temporal sampling schemes. The
acquisition information must remain associated with the measurement
because it determines how the recorded decay should be processed and
interpreted.

The analysis landscape is similarly diverse. Reconvolution-based NLSF
and MLE estimate explicit exponential decay parameters
\citep{Kollner1992}. Phasor analysis provides a fit-free representation
of the measured decay \citep{Digman2008}. Laguerre expansion estimates
decay behavior without requiring a fixed multiexponential component
count \citep{Jo2005}, while RLD provides a computationally inexpensive
lifetime estimate. These approaches are complementary and can provide
different information from the same measurement, making direct
comparison on identical data valuable.

Method development introduces a related need for realistic data with
known ground truth. Experimental measurements are essential for
evaluating performance under real acquisition conditions, but they
generally do not provide exact per-pixel lifetime parameters. Simulation
can provide those known parameters, provided that the simulated data
reflect the timing, IRF, photon statistics, and detector characteristics
of the acquisition being studied \citep{Smith2019,Pandey2024}.

PyFLI addresses these connected needs by handling detector-specific
data loading and acquisition characteristics within the input and
preprocessing layers, while downstream lifetime-estimation,
reconstruction, and visualization tools use consistent interfaces. The
same experimental dataset can therefore be analyzed with several
supported methods, and its timing, IRF, and acquisition settings can
also be used to configure controlled simulations. This supports
experimental researchers who need reproducible processing across FLI
systems and method developers who need standardized data for training,
validation, and benchmarking.

\section{State of the field}

Several open-source packages provide established tools for particular
parts of the fluorescence-lifetime workflow. \texttt{FLUTE}
\citep{Gottlieb2023}, \texttt{PhasorPy} \citep{PhasorPy2024}, and
\texttt{AlliGator} \citep{Michalet2025} provide Phasor-oriented analysis
and visualization, while \texttt{FLIMfit} \citep{Warren2013} provides a
mature environment for fluorescence-lifetime fitting. These packages
are valuable for the analysis tasks on which they focus.

PyFLI was developed for studies that require experimental data handling,
controlled data generation, multiple estimator families,
reconstruction, and quantitative comparison to operate with compatible
data and result conventions. This design is useful when several
estimators must be applied to the same measurement, when analytical and
learning-based methods must be evaluated against the same simulated
ground truth, or when internally and externally processed results must
be compared within one study. Implementing these components in one
framework also allows acquisition information used during experimental
import to inform the corresponding simulation and evaluation steps.
PyFLI therefore complements specialized fitting and Phasor packages by
supporting a continuous workflow from experimental measurement or
simulated ground truth to parameter estimation and comparison.

\section{Software design}

\begin{figure}[htbp]
    \centering
    \includegraphics[width=\textwidth]{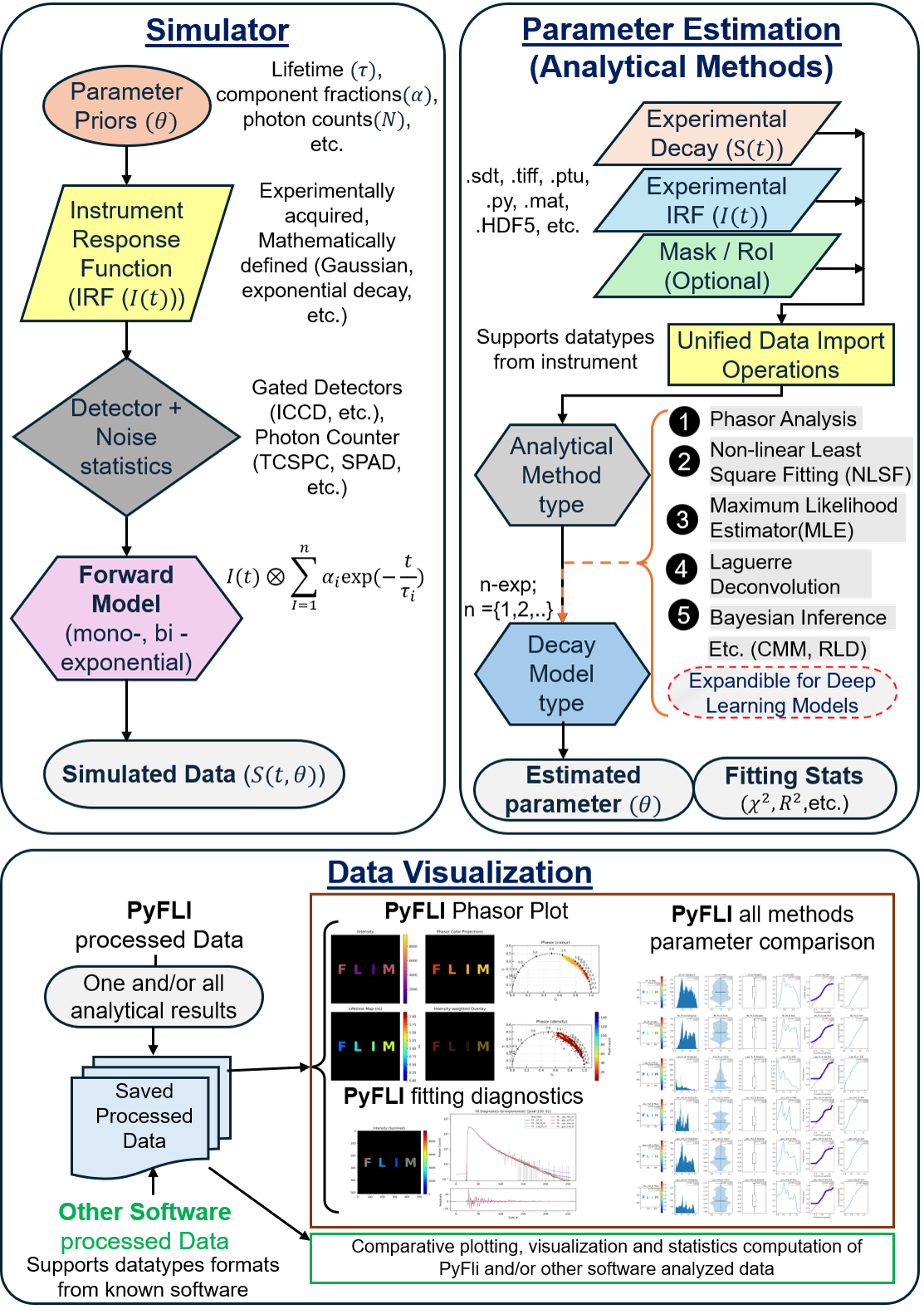}
    \caption{
    Overview of the PyFLI framework. Experimental fluorescence-lifetime
    measurements from supported acquisition systems are imported and
    preprocessed through detector-aware data-handling pathways, while the
    simulation framework generates controlled fluorescence decays with known
    parameters and configurable acquisition and noise characteristics.
    Analytical and trained-inference methods estimate lifetime parameters,
    and common reconstruction, visualization, and comparison tools support
    evaluation across experimental and simulated data.
    }
    \label{fig:pyfli-workflow}
\end{figure}

PyFLI is organized around three related activities: experimental data
handling and parameter estimation, standardized data generation, and
visualization and comparison (Figure~\ref{fig:pyfli-workflow}). Experimental measurements first pass
through readers and preprocessing appropriate to their acquisition
system. PyFLI keeps these acquisition-specific operations separate from
downstream estimation so that analysis methods do not need to be
rewritten for each supported file format or detector. This requires
dedicated readers and metadata handling for each acquisition type, but
it allows the resulting decay data and timing information to enter the
same estimation, simulation, reconstruction, and comparison interfaces.
Estimated parameters and simulated ground truth then use common result
conventions so they can be evaluated with the same downstream tools.
\ifincludesupp
Further architectural details are given in Supplementary
Sections~\ref{supp:architecture} and~\ref{supp:design-principles}.
\fi
Different FLI systems store lifetime information in different forms,
including gated image sequences, photon-arrival histograms, microscopy
containers, and generic numerical arrays. PyFLI imports supported ICCD,
SwissSPAD2/3, and TCSPC measurements together with common scientific
data formats. Supported inputs include \texttt{.sdt}, \texttt{.ptu},
TIFF, NumPy, MATLAB, HDF5, text, and ASC data. The same workflow can
load the associated IRF, background measurements, and masks.
\texttt{.sdt} files are read using \texttt{sdtfile}
\citep{Gohlke2025}, while detector-specific readers assemble gated and
frame-based measurements according to their acquisition structure.
Preprocessing provides background correction, normalization, hot-pixel
handling, region-of-interest operations, and temporal alignment of the
measured decay and IRF. Detector-aware correction and weighting models
account for characteristics including TCSPC pile-up, SPAD counting
behavior, and ICCD gain and read noise.
\ifincludesupp
The input and detector-aware processing paths are described in
Supplementary Sections~\ref{supp:ingestion} and~\ref{supp:deconvolution}.
\fi

Standardized data generation is a central part of the framework because
controlled method evaluation requires known parameters. PyFLI provides
photon-level and continuous forward models that generate fluorescence
decays after convolution with an IRF. Configurable noise processes
include photon-counting noise, dark counts, read noise, gain effects,
quantization, timing jitter, and TCSPC pile-up. Image generators create
spatially heterogeneous lifetime fields and labeled regions, while
batch simulation supports parameter sweeps and training-data generation.
Each simulated dataset retains the parameters used to generate it, so
multiple methods can be tested using the same lifetimes, component
fractions, photon levels, IRF, and acquisition conditions. Calibration
utilities estimate simulator settings from experimental calibration
measurements, validation utilities compare simulated and experimental
noise statistics, and Fisher-information and Cram\'er--Rao calculations
characterize the precision available under specified photon budgets and
acquisition settings.
\ifincludesupp
Simulation, calibration, validation, and precision calculations are
described in Supplementary Sections~\ref{supp:simulation}
and~\ref{supp:precision}.
\fi

PyFLI provides NLSF, Poisson MLE, Phasor analysis, Laguerre expansion,
and RLD within the same analysis framework. Its Phasor implementation
includes acquisition-aware formulations for binned, gated, truncated,
and temporally offset measurements \citep{Michalet2021}. Image-scale
processing is available on CPU and GPU, with spatial binning, global
fitting strategies, model comparison, and per-pixel Cram\'er--Rao
uncertainty estimates available within the fitting framework. A
reconstruction layer converts fitted parameter maps back into modeled
decay cubes, allowing estimated decays to be compared directly with the
original measurements. An optional Bayesian and deep-learning pathway
applies trained posterior-sampling models to decay images and evaluates
posterior-derived parameter combinations against the measurements.
\ifincludesupp
The analytical estimators, image-scale solvers, reconstruction, and
trained-inference pathway are described in Supplementary
Sections~\ref{supp:estimators}, \ref{supp:solver},
\ref{supp:reconstruction}, and~\ref{supp:bayes}.
\fi

Visualization and analysis tools compare parameter maps, lifetime
distributions, fitting diagnostics, classifications, and statistical
summaries across methods and datasets. Results produced by external FLI
software can enter the same comparison workflow. Region-of-interest
tools support manual and automated masks, and statistical utilities
provide hypothesis tests, effect-size measures, correlation analysis,
and factor analysis. For compressive FLI, PyFLI provides Hadamard and
DCT sensing bases with linear, total-variation, and Poisson-likelihood
reconstruction for single-pixel hyperspectral time-resolved imaging
\citep{Pian2017}. PyFLI requires Python 3.11 or later, is distributed
through PyPI as \texttt{pyfli-lib}, and is documented at
\url{https://pyfli.org}. Automated tests cover the major data,
simulation, estimation, reconstruction, visualization, and
trained-inference workflows, with continuous integration on Ubuntu,
Windows, and macOS.
\ifincludesupp
Additional implementation, usage, and quality-control details are given
in Supplementary Sections~\ref{supp:comparison},
\ref{supp:implementation}, \ref{supp:examples}, and~\ref{supp:qc}.
\fi

\section{Research impact statement}

PyFLI has already been used as research infrastructure in a separate
computational imaging study. Erbas et al.\ used the PyFLI simulation
framework to generate a dataset of 1.6 million simulated
fluorescence-lifetime signals for investigating low-precision recurrent neural-network inference \citep{erbas2026quantization} and also partly in the previous work \citep{pandey2025real}. The simulated signals combined mono- and biexponential fluorescence kinetics, photon statistics, detector-dependent noise, and experimentally measured pixel-wise instrument response functions. PyFLI-generated data provided the common basis for training, validation, and held-out evaluation in that study. 

Together, these research uses illustrate the two roles for which PyFLI was designed: providing standardized experimental-data processing across FLI systems and providing controlled simulated data with known parameters for method development and benchmarking. The same surrounding tools for preprocessing, reconstruction, visualization, and quantitative comparison can therefore be retained as a method moves from controlled simulation to experimental fluorescence-lifetime measurements.
\section{AI usage disclosure}

Large language models were used to add comments and edit the documentation within the codebase. All scientific content, software design decisions, code, and final wording were reviewed and approved by the author. The AI was not used to generate experimental results, software behavior claims, or citations; references were verified by the author.

\section*{Acknowledgements}

We acknowledge Dr$.$~Xavier Michalet's work on Phasor analysis \citep{Michalet2021}. We thank Nanxue Yuan, Saif Ragab, Catherine Sherry, Isaiah Crosbourne, Dr$.$~Amit Verma, Dr$.$~Mohd$.$ Saqib and Dr$.$~Luis Chavez for using PyFLI in their experimental data analysis and publication work and providing experimental data from testing the PyFLI library. 

\section*{Author Contributions}

Dr. Vikas Pandey conceptualized the project, developed the entire codebase, and serves as the lead maintainer. Ismail Erbas is a co-developer, maintainer and performed testing and validation. Dr. Stefan Radev assisted with repository organization, formatting, and standardization. Prof. Xavier Intes and Margarida Barroso provided funding, developmental support and facilitated the acquisition of biological samples in their respective laboratories. 


\ifincludesupp

\clearpage
\appendix

\section*{Appendices}
\addcontentsline{toc}{section}{Appendices}

\counterwithin{figure}{section}
\counterwithin{table}{section}
\numberwithin{equation}{section}


\section{Software architecture}
\label{supp:architecture}

PyFLI is organized as a collection of related subpackages under the
top-level \texttt{pyfli} namespace. The repository separates data input,
preprocessing, estimation, simulation, reconstruction, visualization,
and statistical analysis so that acquisition-specific operations do not
need to be implemented again inside every estimator. The main
subpackages used by the workflows described in the manuscript are:

\begin{itemize}
    \item \texttt{io}: data and IRF loading, detector dispatch, data
    saving, processed-data import, and SPAD-specific input utilities.
    \item \texttt{data\_cc}: IRF alignment, normalization,
    preprocessing, and region-of-interest operations.
    \item \texttt{Phasor}: class-based and functional Phasor analysis,
    including acquisition-aware formulations.
    \item \texttt{laguerre}: Laguerre-expansion lifetime estimation and
    deconvolution.
    \item \texttt{solver}: NLSF and MLE fitting, CPU and GPU image
    processing, binned fitting, global fitting, model comparison, and
    shared derived metrics.
    \item \texttt{irf\_deconvolution}: detector-specific observation
    models and joint decay and IRF deconvolution.
    \item \texttt{simulator}: forward simulation, parameter
    distributions, noise models, image generation, and calibration and
    validation utilities.
    \item \texttt{reconstruction}: reconstruction of decay cubes and
    fit-diagnostic quantities from fitted parameter maps.
    \item \texttt{bayes\_utils}: trained posterior-sampling inference,
    parameter-combination selection, and posterior-fit visualization.
    \item \texttt{sp\_analysis}: sensing bases, measurement simulation,
    and reconstruction for compressive single-pixel FLI.
    \item \texttt{data\_vnp} and \texttt{analysis}: comparative
    visualization, classification, result analysis, hypothesis testing,
    and factor analysis.
    \item \texttt{roi\_maker}: interactive and automated
    region-of-interest tooling.
\end{itemize}

Additional package components provide logging, message display, image
assets, analytical helpers, and single-snapshot workflows. The package
root re-exports selected high-level entry points, while specialized
classes remain available from their owning subpackages.

\section{Design principles}
\label{supp:design-principles}

Several design choices connect the individual subpackages.

First, acquisition-specific logic is kept separate from estimator logic.
Detector readers and detector-specific observation models handle data
organization and measurement characteristics before the decay enters a
lifetime estimator. This permits a common estimator interface to be
used for supported acquisition types without embedding file-format code
inside the fitting classes.

Second, PyFLI uses common decay and result conventions across analytical
methods. The same decay cube, IRF, timing information, and mask can be
passed through several estimators, while fitted parameter maps and
reconstructed decays can be returned to a shared comparison layer. This
reduces the amount of method-specific reshaping required when comparing
results.

Third, CPU and GPU processing follow parallel image-scale workflows. The
CPU and PyTorch-based GPU processors use the same fitting definitions
where supported, while the GPU path provides differentiable parameter
transforms and optional per-pixel precision estimates.

Fourth, simulation and experiment are linked through shared acquisition
parameters and calibration utilities. The simulator can be configured
using timing, IRF, photon-count, and detector settings that correspond
to an experimental acquisition. Calibration and validation classes
provide a way to compare simulated and experimental summary statistics
rather than treating the simulator as a detector-independent source of
training data.

\section{Detector-aware data ingestion}
\label{supp:ingestion}

The input layer supports data from ICCD, SwissSPAD2, SwissSPAD3, and
TCSPC acquisition workflows together with common numerical formats.
Supported inputs include \texttt{.sdt}, \texttt{.ptu}, TIFF, NumPy,
MATLAB, HDF5, text, and ASC data. ICCD measurements can be assembled
from gated image stacks, while SPAD readers handle frame- or
gate-oriented acquisitions. TCSPC data are represented as
photon-arrival histograms or equivalent decay arrays.

The higher-level loading workflow can associate decay data with an IRF,
background measurement, and mask. These objects remain available to
downstream preprocessing and estimation steps. Background correction,
normalization, hot-pixel handling, mask operations, and temporal
alignment can therefore be applied before the same decay is analyzed by
different lifetime estimators.

The \texttt{IRFAligner} class provides global and per-pixel temporal
alignment of the measured decay and IRF. Alignment can use integer
circular shifts or Fourier-domain sub-bin shifts. Keeping IRF alignment
as a separate preprocessing operation allows the same aligned response
to be used by reconvolution fitting, Phasor calibration, reconstruction,
and simulation workflows.

\begin{table}[htbp]
\centering
\caption{Acquisition families and input forms used by the PyFLI loading layer.}
\label{tab:supp-readers}
\begin{tabular}{p{0.20\textwidth}p{0.31\textwidth}p{0.39\textwidth}}
\hline
Acquisition family & Representative input & Processing context \\
\hline
ICCD & TIFF gate stacks & Wide-field gated decay measurements with
intensifier gain and read-noise considerations. \\
SwissSPAD2/3 & HDF5 and supported native frame or gate data & Gated
single-photon measurements with detector-specific counting behavior. \\
TCSPC & \texttt{.sdt}, \texttt{.ptu}, and exported numerical arrays &
Photon-arrival histograms with acquisition-dependent pile-up behavior. \\
Generic arrays & TIFF, NumPy, MATLAB, HDF5, text, ASC & Neutral array
input for data already exported from an acquisition-specific workflow. \\
\hline
\end{tabular}
\end{table}

\section{Analytical lifetime estimation}
\label{supp:estimators}

FLI decays can be modeled as sums of exponential components convolved
with an instrument response. For a decay sampled at times $t_k$ with
$m$ components, the model used by reconvolution fitting can be written
as

\begin{equation}
\hat{y}(t) =
b + \mathrm{IRF}_{s}(t) *
\sum_{i=1}^{m} A_i
\exp\left(-\frac{t}{\tau_i}\right),
\label{eq:supp-forward-model}
\end{equation}

where $A_i$ and $\tau_i$ are the amplitude and lifetime of component
$i$, $\mathrm{IRF}_{s}$ is the temporally aligned instrument response,
$*$ denotes convolution, and $b$ is a background term when included by
the selected fitting model. Amplitude fractions and the
amplitude-weighted mean lifetime are

\begin{equation}
\alpha_i = \frac{A_i}{\sum_j A_j},
\qquad
\tau_m = \sum_i \alpha_i \tau_i.
\label{eq:supp-derived}
\end{equation}

PyFLI provides several complementary approaches to estimating lifetime
quantities from the same decay data.

\subsection{Nonlinear least-squares fitting}

The NLSF path estimates explicit model parameters by minimizing a
weighted residual objective between the measured decay and the
reconvolved model,

\begin{equation}
\chi^2(\theta) =
\sum_k w_k
\left(y_k-\hat{y}_k(\theta)\right)^2,
\label{eq:supp-nlsf}
\end{equation}

where $w_k$ denotes the selected weighting scheme. The fitting
framework supports per-pixel and image-scale analysis, configurable fit
ranges, monoexponential and biexponential models, derived lifetime
metrics, and model-comparison utilities.

\subsection{Maximum-likelihood estimation}

For photon-counting data, PyFLI also provides a Poisson
maximum-likelihood objective. Ignoring terms that are constant with
respect to the model parameters, the negative log-likelihood can be
written as

\begin{equation}
-\log L(\theta) =
\sum_k
\left[
\hat{y}_k(\theta)
-
y_k \log \hat{y}_k(\theta)
\right].
\label{eq:supp-mle}
\end{equation}

NLSF and MLE use the same surrounding image-processing and
model-comparison framework, which allows both objectives to be applied
to the same decay data.

\subsection{Phasor analysis}

Phasor analysis maps each decay to Fourier coordinates rather than
fitting an explicit exponential model \citep{Digman2008}. For harmonic
$n$ and angular frequency $\omega$, the standard discrete transform is

\begin{equation}
g(n) =
\frac{\sum_k y_k \cos(n\omega t_k)}
     {\sum_k y_k},
\qquad
s(n) =
\frac{\sum_k y_k \sin(n\omega t_k)}
     {\sum_k y_k}.
\label{eq:supp-phasor}
\end{equation}

The class-based \texttt{PhasorAnalyzer} supports image-scale Phasor
computation and IRF-based calibration. A complementary functional API
implements acquisition-aware single-exponential Phasor-locus
formulations for continuous, binned, gated, truncated, and temporally
offset measurements following \citep{Michalet2021}. These
formulations are relevant when a finite gate width or recording window
changes the Phasor geometry relative to an ideal continuous decay.

\subsection{Laguerre expansion}

The Laguerre pathway represents decay behavior using an orthonormal
Laguerre basis rather than assuming a fixed multiexponential form at
the deconvolution stage \citep{Jo2005}. The implementation constructs
the basis, estimates expansion coefficients, reconstructs the impulse
response, and derives lifetime-related quantities from the result.
Optional coefficient constraints and regularization are available in
the fitting workflow.

\subsection{Rapid lifetime determination}

RLD provides a low-cost lifetime estimate from ratios of integrated
decay regions. In PyFLI, RLD-related calculations can be used directly
or as part of an initial-estimation pathway for iterative fitting.
This provides a rapid estimate when a full nonlinear fit is not
required for every intermediate processing step.

\begin{table}[htbp]
\centering
\caption{Complementary lifetime-estimation pathways in PyFLI.}
\label{tab:supp-estimators}
\begin{tabular}{p{0.16\textwidth}p{0.28\textwidth}p{0.46\textwidth}}
\hline
Estimator & Main assumption & Role in the analysis workflow \\
\hline
NLSF & Fixed exponential model convolved with an IRF &
Explicit lifetime and amplitude estimation using a residual objective. \\
MLE & Fixed exponential model with a photon-counting likelihood &
Explicit lifetime and amplitude estimation under a Poisson objective. \\
Phasor & Fourier representation of the measured decay &
Fit-free visualization and lifetime-related analysis, including
acquisition-aware Phasor geometry. \\
Laguerre & Basis expansion of the decay response &
Deconvolution without fixing the exponential component count at the
basis-expansion stage. \\
RLD & Ratios of integrated decay regions &
Low-cost lifetime estimation and initialization. \\
\hline
\end{tabular}
\end{table}

\section{Detector-specific IRF deconvolution}
\label{supp:deconvolution}

The \texttt{irf\_deconvolution} subpackage represents detector-specific
measurement statistics explicitly rather than applying one observation
model to all acquisitions. It provides detector parameter classes and
observation functions for TCSPC, SPAD, and ICCD measurements together
with weighting and regularized deconvolution utilities.

For TCSPC data, the observation model includes photon-counting
statistics and an optional pile-up transformation that depends on the
excitation count. For gated SPAD data, the model represents repeated
single-photon detection using a binomial observation process. For ICCD
data, the model includes gain and read-noise terms associated with the
intensified detector. These detector-specific transformations and
weights are used before or within deconvolution rather than being
implicitly absorbed into a universal least-squares model.

The joint deconvolution workflow also provides spatial regularization
options. These operations are separate from the simpler loading and
preprocessing corrections in the \texttt{io} and \texttt{data\_cc}
subpackages.

\section{Image-scale solver framework}
\label{supp:solver}

The \texttt{solver} subpackage provides CPU and PyTorch-based GPU image
processors around the common fitting definitions. The GPU path exposes
NLSF and MLE through the image-fitting interface and can optionally
return per-pixel Cram\'er--Rao uncertainty estimates together with the
fitted parameters.

The GPU implementation uses differentiable transformations to enforce
physical parameter constraints during optimization. Positive scalar
parameters can be represented through exponential transforms, bounded
fractions through sigmoid transforms, and ordered lifetime parameters
through offset-based parameterizations. These transformations permit
gradient-based optimization while keeping fitted values within the
required physical domain.

PyFLI also provides spatial binning and global-fitting utilities.
Binning trades spatial resolution for increased photon counts before
fitting. Global-fitting workflows combine information across related
pixels or groups according to the selected configuration. These paths
provide alternatives to completely independent per-pixel fitting in
low-signal or computationally demanding datasets.

\section{Decay reconstruction and fit diagnostics}
\label{supp:reconstruction}

The \texttt{reconstruction} subpackage converts fitted parameter maps
back into modeled decay cubes. \texttt{ParamToDecay} performs the
parameter-to-decay conversion so that the reconvolved model can be
compared directly with the original measurement. This comparison is
downstream of the estimator and can therefore be applied to parameter
maps produced by different fitting pathways.

Additional reconstruction utilities provide monoexponential and
biexponential splitting, dominant-lifetime classification, and
fit-quality maps. These outputs support per-pixel diagnostics and
cross-method comparison without requiring the visualization layer to
reimplement the forward model.

\section{Simulation, calibration, and validation}
\label{supp:simulation}

The simulation layer generates fluorescence decays for which the model
parameters and acquisition settings are known. PyFLI contains
photon-level and continuous simulation pathways, parameter-distribution
samplers, noise models, image generators, and batch workflows. Simulated
decays are generated from model parameters and an IRF, after which
measurement and detector effects can be applied.

Configurable simulation components include photon-counting noise, dark
counts, Gaussian read noise, gain effects, quantization, timing jitter,
and TCSPC pile-up. Higher-level image generators produce spatially
heterogeneous lifetime fields and labeled regions. Batch simulation can
therefore generate datasets across controlled ranges of lifetimes,
component fractions, photon levels, and acquisition settings.

The \texttt{FLICalibrator} class estimates simulator settings from an
experimental calibration dataset. The calibration workflow varies
simulator parameters and compares simulated and experimental summary
statistics. \texttt{FLIValidator} provides complementary comparisons of
simulated and experimental behavior. These tools are intended to make
the assumptions used for training-data generation and benchmarking
explicit and testable.

\section{Precision calculations}
\label{supp:precision}

The simulation and fitting layers include Fisher-information and
Cram\'er--Rao calculations for selected lifetime models. For a
parameter vector $\theta$ with Fisher information matrix $I(\theta)$,
the variance of an unbiased estimator is bounded by

\begin{equation}
\mathrm{Var}(\hat{\theta}_i)
\ge
\left[I(\theta)^{-1}\right]_{ii}.
\label{eq:supp-crlb}
\end{equation}

These calculations provide a model-based description of the precision
available under specified photon budgets and acquisition settings. The
GPU fitting workflow can also return per-pixel Cram\'er--Rao estimates
during image fitting. The resulting precision calculations complement
empirical comparisons with simulated ground truth and experimental
data.

\section{Bayesian and deep-learning direct inference}
\label{supp:bayes}

The \texttt{bayes\_utils} subpackage provides an optional
posterior-sampling pathway built around a trained BayesFlow/Keras model.
The \texttt{BiPipeline} interface applies a trained posterior-sampling
model to a decay image, while \texttt{ParamSelector} evaluates
combinations of posterior-derived parameters against the measured
decay. A plotting utility shows the posterior-derived fit for an
individual pixel.

Keras and TensorFlow are installed through the optional
\texttt{pyfli-lib[tf]} dependency group so that the core analytical
package can be used without the trained-inference dependencies. The
pathway assumes that a compatible posterior model has already been
trained for the acquisition and parameterization under study.

\section{Compressive single-pixel imaging}

The \texttt{sp\_analysis} subpackage supports compressive single-pixel
FLI workflows. It includes Hadamard and DCT sensing bases, measurement
simulation, and linear and regularized reconstruction. A
photon-counting reconstruction path is provided for SPAD-like
measurements. These components follow the single-pixel hyperspectral
time-resolved imaging setting described by Pian et al.
\citep{Pian2017}.

\section{Comparison, classification, visualization, and ROI analysis}
\label{supp:comparison}

The \texttt{data\_vnp} and \texttt{analysis} subpackages provide common
visualization and comparison tools for parameter maps, lifetime
distributions, fit diagnostics, classifications, and statistical
summaries. Processed-data importers can load results generated outside
PyFLI into the same comparison layer. This permits internal and external
results to be evaluated with shared plotting and statistical utilities.

The comparison layer also includes utilities for monoexponential and
biexponential classification and for correlations between parameter
maps produced by different methods. Statistical functions include
hypothesis testing, effect-size calculations, correlation analysis, and
factor analysis. These functions operate on arrays supplied by the
analysis workflow; selection of the appropriate biological or
experimental unit remains the responsibility of the user.

Region-based analysis can use externally supplied masks or masks created with the \texttt{roi\_maker} subpackage. The interactive ROI editor allows manual drawing, while automated options support threshold- and
clustering-based masks. Region-level decay aggregation can improve
photon statistics, with the corresponding trade-off that heterogeneity
within the selected region is averaged.

\section{Implementation and availability}
\label{supp:implementation}

PyFLI requires Python 3.11 or later and is distributed on PyPI as
\texttt{pyfli-lib}. The package uses NumPy and SciPy for numerical
operations, scikit-image and OpenCV for image processing, pandas and
scikit-learn for analysis utilities, Matplotlib and related plotting
packages for visualization, PyTorch for tensor and GPU computation,
\texttt{sdtfile} and \texttt{tifffile} for supported scientific data
input, and PySide6 for the interactive ROI editor.

The core installation includes the analytical and PyTorch components.
The optional \texttt{gpu} dependency group installs CUDA runtime
packages used by accelerated workflows. The optional \texttt{tf}
dependency group installs TensorFlow and Keras for the
posterior-sampling pathway. Documentation is provided through the
project site at \url{https://pyfli.org}.

\section{Example usage}
\label{supp:examples}

A typical workflow loads decay data and a matched IRF, performs temporal
alignment, and applies an image-scale fit:

\begin{verbatim}
from pyfli import DataOperations
from pyfli.data_cc import IRFAligner
from pyfli.solver import BaseFLIFitter, FLIGPUProcessor
loader = DataOperations(
    data_path="experiment.sdt",
    irf_path="irf.txt",
    bg_path="background.tif",
    mask_path="mask.png",
)
decay = loader.load_data(sub_bg=True, hot_pixel=True)
irf = loader.load_irf()
irf_aligned = IRFAligner(decay, irf).align()
processor = FLIGPUProcessor(
    freq = 80,
    fitter_class=BaseFLIFitter,
)
result = processor.fit_image(
    decay,
    irf_aligned,
    mode="NLSF",
    model_type="bi-exponential",
    CRLB=True,
)
\end{verbatim}

A Phasor analysis can be run on the same decay data:

\begin{verbatim}
from pyfli.phasor.phasorS import PhasorAnalyzer
phasor = PhasorAnalyzer(
    frequency_hz=80e6,
    time_axis_ns=time_axis,
    n_harmonics=1,
)
g, s = phasor.create_phasor_cpu(decay)
g_cal, s_cal = phasor.calibrate(g, s, irf_aligned)
\end{verbatim}

For simulation-based method development, the image generator can produce
synthetic decay data and corresponding parameter maps:

\begin{verbatim}
from pyfli import FLIModelImageGenerator
generator = FLIModelImageGenerator(
    irf_data=irf,
    image_shape=(64, 64),
    method="ICCD",
)
synthetic = generator.generate_image()
\end{verbatim}

The examples above are intended to show the relationship between the
input, estimation, and simulation layers. Complete and version-specific
usage examples are maintained in the package documentation.

\section{Quality control}
\label{supp:qc}

The repository contains a pytest test suite covering analytical and
image-scale fitting, the GPU solver, binned fitting, fitting comparison,
Phasor analysis, Laguerre estimation, noise models, IRF alignment, SPAD
input, decay reconstruction, simulation, Bayesian-inference utilities,
data saving, visualization, and analysis components. Tests use
self-contained synthetic arrays where possible.

Continuous integration runs the test suite on Ubuntu, Windows, and
macOS with Python 3.11. Separate workflows check code style and
documentation. The simulator also includes calibration and validation
utilities for comparing simulated behavior with experimental
calibration data. These checks complement direct cross-method
comparisons on shared decay data.

\section{Scope and limitations}

Detector-specific corrections and simulations depend on acquisition
parameters supplied by the user or estimated from calibration data.
Incorrect detector settings can therefore change the resulting
correction or simulated distribution. Full per-pixel fitting of large
images can remain computationally demanding despite GPU execution,
binning, and global-fitting alternatives.

The library is programmatic and assumes familiarity with Python.
Detector coverage is tied to implemented and tested readers rather than
to every file type produced by a detector family. The
posterior-sampling pathway likewise requires a trained model compatible
with the acquisition and parameterization under study.

Time-lapse object tracking and joint multi-dataset modeling beyond the
current comparison layer are not part of the present workflow. Broader
benchmarking against external software on experimental datasets also
depends on appropriate reference datasets and matched processing
configurations. These limitations do not change the common data,
simulation, estimation, and comparison interfaces described in the main
text.

\fi


\clearpage
\bibliographystyle{unsrtnat}
\bibliography{refs}

@article{digman2008,
  title={The phasor approach to fluorescence lifetime imaging analysis},
  author={Digman, Michelle A and Caiolfa, Valeria R and Zamai, Moreno and Gratton, Enrico},
  journal={Biophysical journal},
  volume={94},
  number={2},
  pages={L14--L16},
  year={2008},
  publisher={Elsevier}
}

@article{jo2005,
  title={Ultrafast method for the analysis of fluorescence lifetime imaging microscopy data based on the Laguerre expansion technique},
  author={Jo, Javier A and Fang, Qiyin and Marcu, Laura},
  journal={IEEE Journal of Selected Topics in Quantum Electronics},
  volume={11},
  number={4},
  pages={835--845},
  year={2005},
  publisher={IEEE}
}

@article{bruschini2019,
  title={Single-photon avalanche diode imagers in biophotonics: review and outlook},
  author={Bruschini, Claudio and Homulle, Harald and Antolovic, Ivan Michel and Burri, Samuel and Charbon, Edoardo},
  journal={Light: Science \& Applications},
  volume={8},
  number={1},
  pages={87},
  year={2019},
  publisher={Nature Publishing Group UK London}
}

@article{smith2019,
  title={Fast fit-free analysis of fluorescence lifetime imaging via deep learning},
  author={Smith, Jason T and Yao, Ruoyang and Sinsuebphon, Nattawut and Rudkouskaya, Alena and Un, Nathan and Mazurkiewicz, Joseph and Barroso, Margarida and Yan, Pingkun and Intes, Xavier},
  journal={Proceedings of the national academy of sciences},
  volume={116},
  number={48},
  pages={24019--24030},
  year={2019},
  publisher={National Academy of Sciences}
}

@article{pian2017,
  title={Compressive hyperspectral time-resolved wide-field fluorescence lifetime imaging},
  author={Pian, Qi and Yao, Ruoyang and Sinsuebphon, Nattawut and Intes, Xavier},
  journal={Nature photonics},
  volume={11},
  number={7},
  pages={411--414},
  year={2017},
  publisher={Nature Publishing Group UK London}
}

@article{kollner1992,
  title={How many photons are necessary for fluorescence-lifetime measurements?},
  author={K{\"o}llner, Malte and Wolfrum, J{\"u}rgen},
  journal={Chemical Physics Letters},
  volume={200},
  number={1-2},
  pages={199--204},
  year={1992},
  publisher={Elsevier}
}

@article{warren2013,
  title={Rapid global fitting of large fluorescence lifetime imaging microscopy datasets},
  author={Warren, Sean C and Margineanu, Anca and Alibhai, Dominic and Kelly, Douglas J and Talbot, Clifford and Alexandrov, Yuriy and Munro, Ian and Katan, Matilda and Dunsby, Chris and French, Paul MW},
  journal={PloS one},
  volume={8},
  number={8},
  pages={e70687},
  year={2013},
  publisher={Public Library of Science San Francisco, USA}
}

@article{michalet2021,
  title={An overview of continuous and discrete phasor analysis of binned or time-gated periodic decays},
  author={Michalet, Xavier},
  journal={Multiphoton Microscopy in the Biomedical Sciences XXI},
  volume={11648},
  pages={34--45},
  year={2021},
  publisher={SPIE}
}

@article{gottlieb2023,
  title={FLUTE: A Python GUI for interactive phasor analysis of FLIM data},
  author={Gottlieb, Dale and Asadipour, Bahar and Kostina, Polina and Ung, Thi Phuong Lien and Stringari, Chiara},
  journal={Biological imaging},
  volume={3},
  pages={e21},
  year={2023},
  publisher={Cambridge University Press}
}

@misc{PhasorPy2024,
  author       = {{PhasorPy contributors}},
  title        = {{PhasorPy: An open-source Python library for phasor analysis of fluorescence lifetime and hyperspectral imaging}},
  year         = {2024},
  howpublished = {\url{https://www.phasorpy.org/}},
  note         = {Software package}
}

@misc{Gohlke2025,
  author       = {Gohlke, Christoph},
  title        = {{sdtfile: Read Becker \& Hickl SDT files}},
  year         = {2025},
  howpublished = {\url{https://github.com/cgohlke/sdtfile}},
  doi          = {10.5281/zenodo.10125608},
  note         = {Software package}
}

@article{pandey2024,
  title={Deep learning-based temporal deconvolution for photon time-of-flight distribution retrieval},
  author={Pandey, Vikas and Erbas, Ismail and Michalet, Xavier and Ulku, Arin and Bruschini, Claudio and Charbon, Edoardo and Barroso, Margarida and Intes, Xavier},
  journal={Optics letters},
  volume={49},
  number={22},
  pages={6457--6460},
  year={2024},
  publisher={Optica Publishing Group}
}

@article{michalet2025,
  title={AlliGator: Open source fluorescence lifetime imaging analysis in G},
  author={Michalet, Xavier},
  journal={SoftwareX},
  volume={31},
  pages={102255},
  year={2025},
  publisher={Elsevier}
}

@misc{erbas2026quantization,
      title={When Quantization Breaks Memory: Recurrent-State Write-Back in Low-Precision Temporal Inference}, 
      author={Ismail Erbas and Xavier Intes and Vikas Pandey},
      year={2026},
      eprint={2609.04490},
      archivePrefix={arXiv},
      primaryClass={cs.AI},
      url={https://arxiv.org/abs/2609.04490}, 
}

@article{pandey2025real,
  title={Real-time wide-field fluorescence lifetime imaging via single-snapshot acquisition for biomedical applications},
  author={Pandey, Vikas and Millar, Euan and Erbas, Ismail and Chavez, Luis and Radford, Jack and Crosbourne, Isaiah and Madhusudan, Mansa and Taylor, Gregor G and Yuan, Nanxue and Bruschini, Claudio and others},
  journal={PhotoniX},
  volume={6},
  number={1},
  pages={58},
  year={2025},
  publisher={Springer}
}

\end{document}